\documentclass{elsart}

\usepackage{amsmath}
\usepackage{amsfonts}
\usepackage{amssymb}
\usepackage{graphicx}
\usepackage{natbib}
\usepackage{multirow}
\usepackage{rotating}

\newtheorem{theorem}{Theorem}
\newtheorem{proof}{Proof}

\newcommand{\cinlaw}{\buildrel  { \mathcal {L} }\over{\longrightarrow}}
\newcommand{\cinP}{\buildrel  { \mathcal{P} }\over{\longrightarrow}}
\newcommand{\Prof}{{\mathbf P}}
\newcommand{\hProf}{\mathbf {\widehat {\Prof}}}
\newcommand{\eProf}{{\mathbf{\mathcal{P}}}}
\newcommand{\heProf}{\mathbf{\widehat{\eProf}}}
\newcommand{\nmu}{n\mathbf{\mu}}
\newcommand{\nSigma}{n\mathbf{\Sigma}}
\newcommand{\rank}{rank}
\newcommand{\dPPO}{d(\Prof,\Prof_0)}
\newcommand{\dhPPO}{d(\hat \Prof,\Prof_0)}

\newcommand{\chiCL}{\overline{\chi}_\beta^2}

\newcommand{\R}{\texttt{R}}

\newcommand{\Rpackage}[1]{{\textit{#1}}}

\begin{document}

\begin{frontmatter}
\title{Distance based Inference for \\ Gene-Ontology Analysis of \\
Microarray Experiments}

\author{Alex S\'anchez-Pla\corauthref{cor}}
\address{Departament d'Estad\'istica. Facultat de Biologia.\\
Avda Diagonal 645. 08028 Barcelona, Spain} \ead{asanchez@ub.edu}

\author{Miquel Salicr\'u}
\address{Departament d'Estad\'istica. Facultat de Biologia.\\
Avda Diagonal 645. 08028 Barcelona, Spain}
\ead{msalicru@ub.edu}

\author{Jordi Oca\~na}
\address{Departament d'Estad\'istica. Facultat de Biologia.\\
Avda Diagonal 645. 08028 Barcelona, Spain}
\ead{jocana@ub.edu}

\corauth[cor]{Alex S\'anchez}
\begin{abstract}
The increasing availability of high throughput data arising from
gene expression studies leads to the necessity of methods for
summarizing the available information. As annotation quality
improves it is becoming common to rely on the Gene Ontology (GO)
to build \emph{functional profiles} that characterize a set of
genes using the frequency of use of each GO term or group of terms
in the array. In this work we describe a statistical model for
such profiles, provide methods to compare profiles and develop
inferential procedures to assess this comparison. An {\tt
R}-package implementing the methods is available.
\end{abstract}


\end{frontmatter}

\section{Introduction}
DNA microarrays belong to recently developed technologies which
allow to measure the expression of thousands of genes
simultaneously, in a single experiment. It is expected that these
experiments will contribute to solve many relevant biological
problems ranging from the identification of complex genetic
diseases, \cite{Alon:1999}, or the prediction of tumor type,
\cite{Alizadeh:2000}, to target discovery the pharmaceutical
industry.

A common trait in these type of studies is the fact that they
generate huge quantities of data and one may end with lists of
up-to thousands of genes which need to be given a biological
interpretation.

A typical microarray experiment is one who looks for genes
\emph{differentially expressed} between two or more conditions.
That is, genes which behave differently in one condition (for
instance healthy [or untreated or wild-type] cells) than in
another (for instance tumour [or treated or mutant] cells). The
study by \cite{Hengel:2003} is of this type and will be used as an
illustration of the ideas discussed in this paper. These authors
showed that memory $CD4+$ T--cells behave differently if they
present ($CD62L+$) or they lack ($CD62L-$) expression. In a study
oriented to find the genetic regulation of these differences they
found 144 genes to be upregulated in $CD4+/CD62L-$ T--cells
relative to $CD4+/CD62L+$ T--cells. Methods such as those
presented here have been developed to contribute to the biological
interpretation of the resulting lists of genes. To do this they
rely on the Gene Ontology, which is described in the following.

\subsection{The Gene Ontology}

\begin{figure}
\begin{center}
\includegraphics[width=13cm]{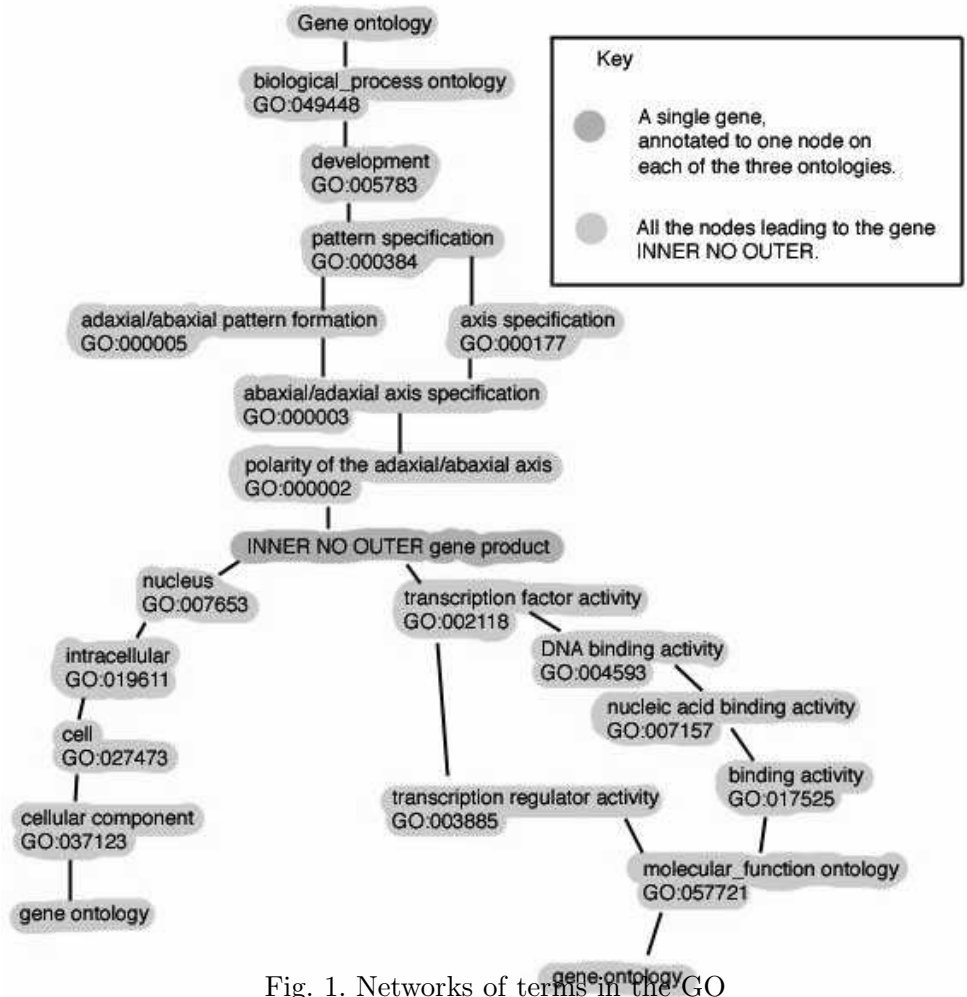}
\vspace{-1cm} \caption{Networks of terms in the GO} \label{GO.nodes}
\end{center}
\end{figure}

Attempts to perform a biological interpretation of these
experiments are often based on the Gene Ontology (GO), an
annotation database created and maintained by a public consortium,
the Gene Ontology Consortium \footnote{www.geneontology.org},
whose main goal is, citing their mission, \emph{to produce a
controlled vocabulary that can be applied to all organisms even as
knowledge of gene and protein roles in cells is accumulating and
changing}. The GO is organized around three principles or basic
ontologies: (1) Molecular function (MF), which describes tasks
performed by individual gene products such as transcription factor
or ATPase activity; (2) Biological process (BP), which describes
broad biological goals, such as mitosis or purine metabolism, that
are accomplished by ordered assemblies of molecular functions, and
(3) Cellular component (CC) describing subcellular structures,
locations, and macromolecular complexes such as nucleus, telomere,
and origin recognition complex. A given gene product may represent
one or more molecular functions, be used in one or more biological
processes and appear in one or more cellular components (see
Figure \ref{GO.nodes}).

Figure ~\ref{GO.nodes} shows how a given gene product can be
characterized using different related terms in each ontology. An
important point to note is that the information here is not
``linear'' in the sense that, although there is a hierarchical
relation, there are interrelations between levels and terms in
each ontology. As a consequence an appropriate representation for
this figure is a \emph{directed acyclical graph (DAG)} and several
analysis methods rely precisely on this representation. This is
not the case of the methods discussed here.

\subsection{Summarizing and interpreting microarray experimental results}

The GO is a rich data structure which contains a great quantity of
information about the relation between terms. But due precisely to
this, it is difficult to work with them, all at once. This fact
has motivated that, in recent years, different approaches to
GO--based analysis of the results of high throughput experiments
have been considered. As a result of this effort many methods and
even more tools have been developed. \cite{Mosquera:2005} is a
review of the existing tools, jointly with the questions they try
to solve.

\begin{figure}[htbp]
\includegraphics[width=\textwidth]{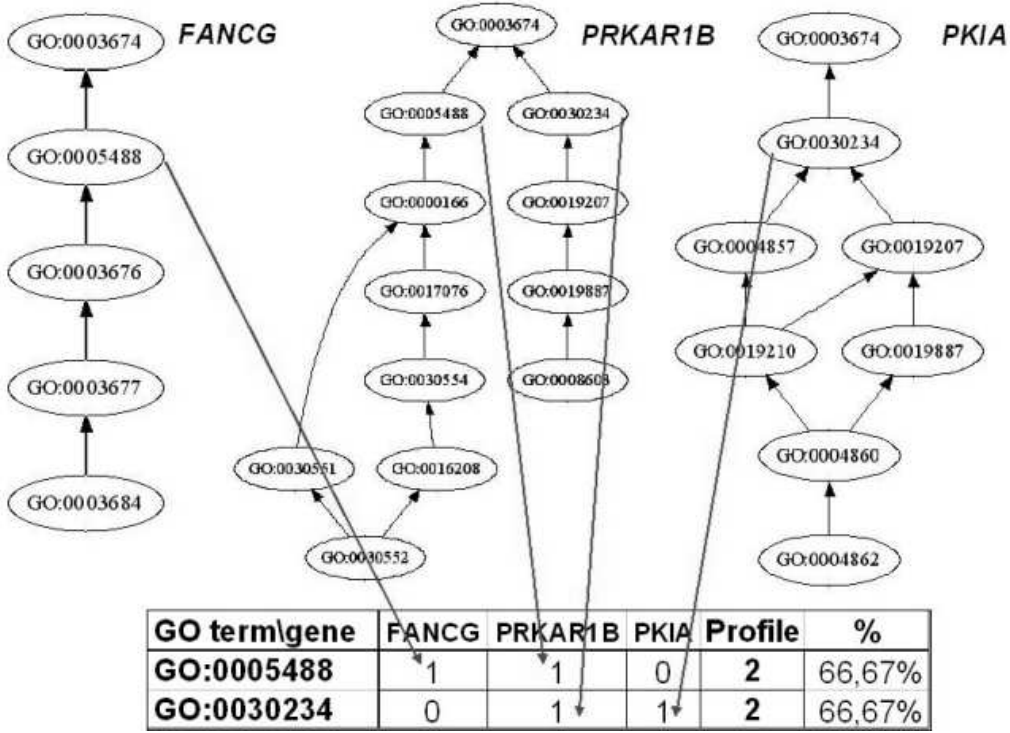}
\caption{\label{simpleProfile}A simple functional profile, based
only on 3 genes, showing the fact that a given gene may appear in
different categories}
\end{figure}

Typically, after having selected one list of interesting genes one
can obtain the \emph{induced sub--graph}, that is the graph formed
by the subset of the Gene Ontology whose nodes are related to the
genes in the list, either directly or through other nodes. These
graphs may be big, complex, structures, specially when the list
originating them is also big. In order to simplify this structure
it may be \emph{sliced} or projected on the nodes which are at a
certain distance of the top node (what is called a \emph{level} of
the GO). This will originate a table of frequencies, with each
cell containing the number of genes annotated by its corresponding
term at the level where the slice has been done (see Figure
\ref{simpleProfile}).
The lattice structure of the graphs implies that one gene may
appear in multiple cells of this table, which we call, from now
on, a \emph{functional profile}. Once this classification has been
done there are different ways to analyze it which are briefly
presented below.
\begin{itemize}
\item One straightforward possibility
is to perform some type of \emph{enrichment analysis} which
consists of a comparison in order to establish if the percentage
of genes in a certain GO category has increased or decreased in
the genes selected relatively to those in the population. If this
is so, a biological explanation of the differences can be
attempted based on this enrichment. This is usually done by means
of a Fisher test or any of its variations and is performed
category-wise for each of the groups in the level selected
followed by some type of multiple testing adjustment.\\
 Programs such as \texttt{fatiGO} (\cite{fatiGO:2004}), \texttt{DAVID}
(\cite{DAVID:2003}) and many more (see \cite{Mosquera:2005})
perform some form of this enrichment analysis. This is by far the
most used approach nowadays.
\item The main characteristic of the previous approach is the fact
that each category is compared separately. A reasonable
complementary extension to this may be to consider all categories
at once and to compare, for instance, the categorization of
selected genes with that of all the genes in the array. There
exist some tools performing this type of comparisons, such as
\texttt{goTools}, a Bioconductor package (\cite{Paquet:2005})
taking this point of view but allowing only for visual
comparisons.
\item Recent works are developing tests which can also be applied to analyze
a whole set of genes selected \emph{a priori} (see
~\cite{Mootha:2003} or ~\cite{Smyth:2005}). Although they are
related in spirit with the previous approaches, the way they
proceed is totally different and will not be discussed here. A
comparison seems however interesting and will be presented
elsewhere.
\end{itemize}

This work is devoted to the modelling and analysis of functional
profiles adopting the intermediate position just described, that
is looking at the profile as a whole more than as a set of
unrelated categories. It will be shown that functional profiles
characterize the set of selected genes and that they can be used
to perform interesting comparisons such as over-expressed vs.
under-expressed genes, or between arrays of different brands.


\section{Statistical models}

A functional profile can be seen as a numerical vector with named
cells. Each cell corresponds to a category in a given ontology,
usually, but not necessarily, at the same level in the GO. Each
category can be characterized by a unique number (GO:nnnnnnn) and a
descriptive name. Saying that a node is at level $k$ means that the
shortest path between this and the main node in each ontology (MF,
BP or CC) has $k-1$ nodes. The cell number represents the number of
genes whose path to the base level has a node in this category.

 Table \ref{TabProfiles} shows a functional profile for a
set of 140 genes clasified at the second level of the Molecular
Function Ontology.

An important thing to have in mind when one considers analyzing
data starting from profiles like that in Table ~\ref{TabProfiles}
is that building this table suppresses the structure of the
original data, as any categorization does and, as a result of the
possibility of a gene to belong to more than one category, the sum
of counts is higher than the number of genes, and in consequence a
different model than the usual multinomial one is needed to
represent cell counts.

\section{The profile distribution}

In order to overpass the problem that a multinomial model is not
adequate to represent a functional profile, the following strategy
is adopted. Given a profile with $s$ categories ($s=8$ in table
~\ref{TabProfiles}) let $\Omega=\{A_1,...,A_s\}$ be the space of
events corresponding to observing one individual in one of the
categories $1$,...,$s$. Given that it is possible that the same
gene belongs to multiple categories we must consider, instead, the
space of events
$$\Omega^*=\{A_1,A_1\times A_2,\dots,A_1\times A_s,A_2,A_2\times
A_3,\dots,A_{s-1}\times A_s\},$$where $A_i$ means that a gene has
appeared only in category $i$ and $A_i \times A_j$ means that it
has appeared in both $i$ and $j$. For simplicity we make our
development assuming that the multiplicity is only for two
categories, but it is straightforward to extend it to more than
two.

Taking this crossed-structure approach, each gene will appear only
once, at most, in each category so that a given experimental
situation may be characterized by an \emph{expanded profile}
\begin{equation}
n\eProf = n(p_{11},p_{12},...,p_{1s},...,p_{(s-1)s},p_{ss})^t
\end{equation}
so that the sample expanded profile, $n\heProf$, is associated
with a multinomial distribution:
\begin{equation}
n\heProf\sim {\mathbf M}(n;\eProf) \label{expansion}
\end{equation}
where $n$ is the number of genes forming the profile (that is,
classified at a given level of a given ontology), $p_{ii}$ is the
probability of a gene to belong only to class $A_i$,
$p_{ij}=p_{ji}$ is the probability of that gene belonging
simultaneously to classes $A_i$ and $A_j$ and $\hat p_{ii}$ and
$\hat p_{ij}$ are the corresponding realizations from a sample of
size $n$.

In practice we are interested in the distribution of the
\emph{contracted} variable $n\widehat\Prof=(n\hat p_{1.},n\hat
p_{2.},...,n\hat p_{s.})^t$ which represents the profile, that is,
the counts in categories $(A_{1.},A_{2.},\dots,A_{s.})$, where
$$
A_{i.}=A_i\bigcup \left(\bigcup_{j<i} A_{j}\times A_{i}\right)
\bigcup \left(\bigcup_{j> i} A_{i}\times A_{j}\right)
$$ and
$p_{i.}$ represents the probability of $A_{i.}$. The distribution
of $n\widehat\Prof$ is established in the following theorem:
\begin{theorem}\label{theorem1}
The random variable ${n\widehat\Prof}=(n\hat p_{1.}, n\hat
p_{2.},\dots,n\hat p_{s.})^t$ is asymptotically distributed as a
multivariate normal distribution
\begin{displaymath}
{\mathbf N}(\mbox{$\nmu$}, \nSigma),
\end{displaymath} where:
$\mathbb{\mu}=(p_{1.},p_{2.},\dots,p_{s.})^t$, and
$\Sigma=(\sigma_{ij})$, with:
 $$
\sigma_{ii}=p_{i.}(1-p_{i.}),\ i=1,...,s,\
\sigma_{ij}=p_{ij}-p_{i.}p_{j.}, i\neq j, i,j=1,...,s.
$$
\end{theorem}
\begin{proof}
The asymptotic normality of ${n\widehat\Prof}$ follows from
considering the asymptotic approximation of the multinomial law to
the normal distribution and the distributional invariance when a
linear transformation is applied to normal distributions (see
\cite{Serfling:1980}), where the transformation is described by:
\begin{equation}\label{contraction}
\hProf =C\heProf
\end{equation}
and $C=\left(C^{(1)}|C^{(2)}|\dots |C^{(s)} \right)$ is a matrix
with $s$ rows and $s\cdot(s+1)/2$ columns defined, by boxes, as:
\begin{equation}
C=\left( \underbrace{\begin{array}{ccccc}
1 & 1 & 1 & \dots & 1 \\
0 & 1 & 0 & \dots & 0 \\
0 & 0 & 1 & \dots & 0 \\
\vdots & \vdots & \vdots & \ddots & \vdots \\
0 & 0 & 0 & \dots & 1\\
\end{array}}_{C^{(1)}}
\left| \underbrace{\begin{array}{ccccc}
0 & 0 & 0 & \dots & 0 \\
1 & 1 & 1 & \dots & 1 \\
0 & 1 & 0 & \dots & 0 \\
\vdots & \vdots & \vdots & \ddots & \vdots \\
0 & 0 & 0 & \dots & 1\\
\end{array}}_{C^{(2)}} \right|
\cdots\left| \underbrace{\begin{array}{c}
  0 \\
  0 \\
  0 \\
  \vdots \\
  1 \\
\end{array}}_{C^{(s)}}\right.
\right)
\end{equation}
with
$$
C^{(h)}=\left (c_{ij}^h\right),
$$
where:
\begin{equation}
c_{ij}^h = \left \{ \begin{array}{ll} 1 & \mbox{if }\ i = h \mbox{ or } (j=i-h+1 \mbox{ and } i \geq h+1)\\
0 & \mbox{ elsewhere}.
\end{array}\right.
\end{equation}
\end{proof}

In the following we will use the term ``profile'' indistinctly to
refer to the absolute frequencies or to the pair formed by the
relative frequencies and the number of genes, $n$.

\section{Comparison of profiles}
In many practical applications the user is interested in comparing
profiles. This is meaningful in a variety of situations, let it be
to compare the profile obtained from a set of over or
under--expressed genes with all the genes on the array, to compare
the profiles obtained in different experimental conditions or to
compare the profiles corresponding to arrays of different types or
brands.


Our approach is based on defining an appropriate measure of
distance between profiles $d({\mathbf P_i}, {\mathbf P_j})$. This
allows to state the problem of comparing two profiles in terms of
testing the hypothesis $H_0: \, d({\mathbf P_i}, {\mathbf P_j})=0$
vs $H_1: \, d({\mathbf P_i}, {\mathbf P_j})>0$.

The choice of the appropriate distance is often a point for
extensive discussion. Sometimes the underlying statistical model is
relevant for its choice. In other cases the availability, or even
computational feasibility is decisive. Here we will use the squared
Euclidean distance which offers a good balance between ease of
interpretation and properties that can be derived for it.

One may consider different scenarios for working with this problem:
\begin{itemize}
\item One--sample problems consist of comparing an estimated profile with a
fixed one, which makes the role of ``population''. This may be, for
instance, a profile obtained from the whole array or even the genome
of the species if it is available.
\item Two--sample problems consist of comparing two (or more)
estimated profiles, which are obtained from populations which may be
or may not be independent. This may be for instance the case of
profiles formed with the genes selected in two different experiments
about the same disease, or those obtained with the genes selected
from two (or more) different mutations of a given wild type.
\end{itemize}
Only the first case will be discussed in the following. Two sample
problems will be presented elsewhere.

\subsection{Main results}

Let $\Prof_0$ represent a fixed population profile, and $\hat
\Prof$, an estimated profile based on a sample of size $n$. The
squared Euclidean distance between ${\mathbf \Prof_0}$ and
${\mathbf {\hat \Prof}}$ is defined as:
\begin{equation}
d({\mathbf {\hat \Prof},\Prof_0})= \left (\hat
\Prof-\Prof_0\right)^t\left (\hat \Prof-\Prof_0\right)=
\sum_{i=1}^s(\hat p_{i.}-p_{i0})^2.
\end{equation}

Based on Theorem \ref{theorem1}, the distribution of the distances
can be established setting the basis to perform statistical
inference on the estimated profiles.

\begin{theorem}\label{theorem2}
Let $\Prof_0$ represent a fixed population profile, $\hat \Prof$
an estimated profile based on a sample of size $n$ and  $d(\hat
\Prof,\Prof_0)$ the squared Euclidean distance between ${\mathbf
\Prof_0}$ and ${\mathbf {\hat \Prof}}$,
\begin{enumerate}
\item
If $\Prof\neq \Prof_0$ then\\
\begin{equation}
n^{1/2}\left [ d (\hat \Prof,\Prof_0)-d\ (\Prof,\Prof_0)\right]
\cinlaw  {\mathbf Z}\sim {\mathbf
N}(0,{\sigma^2}),\label{theorem21}
\end{equation}
where:
\begin{eqnarray}
\sigma^2& =& 4\,\sum_{i,j=1}^s (p_{i.}-p_{i0})\left
(\delta_{ij}p_{i.}+(1-\delta_{ij})p_{ij}-p_{i.}p_{j.}\right
)(p_{j.}-p_{j0}) \nonumber \\&=&
4(\Prof-\Prof_0)^t\mathbf{\Sigma}(\Prof-\Prof_0),\label{SEDist}
\end{eqnarray}
and ``$\cinlaw$'' stands for ``convergence in distribution''.
\item
If $\Prof=\Prof_0$ then
\begin{equation}
n\,d(\hat \Prof,\Prof_0) \cinlaw {\sum_{i=1}^s}
\beta_i\chi^2_{1,i},\label{LinearCombinationChi2}
\end{equation}
where $\beta_i$ are the eigenvalues of matrix $\Sigma$ defined in
Theorem ~\ref{theorem1} and $\chi^2_{1,i}$ are independent
chi-squared random variables with one degree of freedom.
\end{enumerate}
\end{theorem}
\begin{proof}
Consider the algebraic relation:
\begin{eqnarray}
d(\hat \Prof,\Prof_0)&=&\left (\hat \Prof-\Prof_0\right)^t \left (\hat \Prof-\Prof_0\right)\nonumber\\
&=&\left((\hat \Prof-\Prof)-(\Prof_0-\Prof)\right)^t \left((\hat \Prof-\Prof)-(\Prof_0-\Prof)\right) \nonumber \\
&=& d(\hat \Prof,\Prof)+ d (\Prof,\Prof_0) +
2(\Prof-\Prof_0)^t(\hat\Prof-\Prof). \label{algebraic-relation-1}
\end{eqnarray}
Note also that the asymptotic distribution of $n\,d(\hat
\Prof,\Prof)$ follows from standard results about quadratic forms
(\cite{Dik:1985}):
\begin{equation}
n\ d(\hat \Prof,\Prof)=n\ (\hat \Prof-\Prof)^t \mathbb{I}_s (\hat
\Prof-\Prof) \cinlaw {\sum_{i=1}^s} \beta_i\chi^2_{1,i},
\label{quadratic-form-distribution-1}
\end{equation}
where $\mathbb{I}_s$ is the $s-$dimensional identity matrix and
$\beta_i$ are the eigenvalues of $\mathbb{I}_s\Sigma=\Sigma$.

\flushleft Let $\Prof\neq \Prof_0$. If we take the algebraic
relation (\ref{algebraic-relation-1}), the following holds:
\begin{equation}
n^{1/2} \left[d(\hat \Prof,\Prof_0)-d\left
(\Prof,\Prof_0\right)\right ] = n^{1/2} \ d(\hat \Prof,\Prof)+ 2
n^{1/2} \left(\Prof-\Prof_0\right)^t (\hat \Prof-\Prof).
\end{equation}
Then, as $n^{1/2} d(\hat \Prof,\Prof) \cinP 0$ and following
Theorem \ref{theorem1}, the first part of Theorem \ref{theorem2}
is established.

\flushleft Let $\Prof= \Prof_0$. In that case we simply have that
the first two terms of (\ref{algebraic-relation-1}) become null,
so that
\begin{equation}
n\ d (\hat \Prof,\Prof_0)=n\ d (\hat \Prof,\Prof)=n\ (\hat
\Prof-\Prof)^t \mathbb{I}_s (\hat \Prof-\Prof),
\end{equation}
and the second result follows from
(\ref{quadratic-form-distribution-1}).
\end{proof}

The proof of Theorem \ref{theorem2} is based on some properties of
the squared Euclidean distance. It can be easily extended to other
smooth distance indexes with the only condition that they admit a
Taylor series expansion. In that case the approximation appearing
in (\ref{quadratic-form-distribution-1}) would be based on the
eigenvalues of $H_T\Sigma$ where $H_T$ is the Taylor Hessian of
the distance expansion. It must be noted, however, that with
squared Euclidean distance, error terms depend exclusively on the
convergence of the multinomial  to the normal distribution given
that terms of order greater than two vanish in the Taylor
expansion. In absence of other constrains, such as biological
interpretation, this is an additional argument favoring the use of
this distance in front of other options.

In practical situations it is often difficult to deal with linear
combinations of chi--squared random variables.
\cite{raoscott:1981} suggested to use the following approximation
for the combination introduced in equation
~\ref{LinearCombinationChi2}:
\begin{equation}
{\sum_{i=1}^s} \beta_i\chi^2_{1,i}\simeq
\overline{\beta}\chi^2_{\rank \Sigma}
\end{equation}
where
$$
\overline{\beta}=\frac 1 s \left [\sum_{i=1}^s
p_{io}(1-p_{io})\right],
$$
and $\chi^2_{\rank \Sigma}$ stands for a chi--square random
variable with $\rank \Sigma$ degrees of freedom.

Our simulation results show that the above approximation performs
very poorly in our case. In this work we have used a similar
approximation which we consider to have better adaptability
properties:
\begin{equation}
\sum_{i=1}^s \beta_i\chi^2_{1,i}\simeq a\chi^2_{\rank \Sigma}+ b,
\label{aprox-ab}
\end{equation}
where
\begin{equation}
E\left (\chiCL\right )=a\,E\left (\chi^2_{\rank \Sigma}\right ) +
b \quad \mbox{and} \quad Var\left (\chiCL\right ) = a^2\,Var\left
(\chi^2_{\rank \Sigma}\right )
\end{equation}
giving:
\begin{equation}
a=\sqrt{\sum_{i=1}^s \beta_i^2}\quad \mbox{and} \quad
b=\sum_{i=1}^s\beta_i-\sqrt{s\,\sum_{i=1}^s
\beta_i^2}.\label{Define_a_b}
\end{equation}

\subsection{Applications}

These results make it possible to construct hypothesis tests and
confidence intervals to perform statistical inference on the
profiles.

An approximate confidence interval for the ``true'' distance
$d\left( {{\rm {\bf P}},{\rm {\bf P}}_0 } \right)$, with
approximate confidence level $1 - 2\alpha $ is
\begin{equation}\label{ConfidenceInt}
\left[ {\dhPPO - z_\alpha \frac{\hat {\sigma }}{\sqrt n }\,{\kern
1pt} ,\quad \dhPPO + z_\alpha \frac{\hat {\sigma }}{\sqrt n }}
\right]
\end{equation}
where $\hat {\sigma } = 2\sqrt {\left( {{\rm {\bf \hat {P}}} -
{\rm {\bf P}}_0 } \right)^t\hat {\Sigma }\left( {{\rm {\bf \hat
{P}}} - {\rm {\bf P}}_0 } \right)}$ is the sample standard error
estimator of $n\,\dhPPO$, directly obtained from (\ref{SEDist}),
and $z_\alpha $ stands for the $\alpha$ right quantile of a
standard normal random variable $Z$, $\Pr \left\{ {Z \geqslant
z_\alpha } \right\} = \alpha .$

Consider now the contrast
\begin{equation}\label{oneSampleTest}
\begin{array}{l}
 H_0 :\quad \dPPO = 0 \\
 H_1 :\quad \dPPO > 0 \\
 \end{array}
\end{equation}
(Say, is the set of differentially expressed genes
\textit{enriched}/\textit{impoverished} in some GO categories with
respect to all the genes in the array?) From (\ref{theorem21}) the
rule
\begin{equation}\label{RejectionBychiLinComb}
\mbox{``Reject $H_{0}$ if }P\left( \sum_{i=1}^n \beta_i
\chi_{1i}^2 \geq n \hat d \right )\leq \alpha \mbox{''},
\end{equation}
defines a test of nominal size $\alpha$ where $\hat d$ stands for
the sample squared euclidean distance value.

There exist approximate methods to compute tail probabilities of
linear combinations of independent chi--square random variables.
See \cite{Sheil:1977} for the case of non--negative coefficients
or \cite{Farebrother:1984} for more general algorithms.

These algorithms are computationally complex so that we have taken
a more direct approach based on simulation. It consists of
estimating the probability in (\ref{RejectionBychiLinComb}) by
means of the relative frequency
$$
\frac{\#[Y_j\geq n \hat d]}{m},
$$
where $Y_j,\ j=1,\dots,m$, are independent realizations of
$\sum_{i=1}^n \beta_i \chi_{1i}^2$.

In the examples and simulations described in the next sections, we
have taken $m$ = 10,000.

Similarly, the rule
\begin{equation}\label{RejectionBychirank}
 \mbox{``Reject $H_{0}$ if }\
P\left ( \chi_{\rank\Sigma}^2 \geq \frac{n \hat d-b}{a}\right
)\leq \alpha\mbox{''},
\end{equation}
defines an alternative test procedure. Here $a$ and $b$ are
defined in (\ref{Define_a_b}). This last method is slightly easier
to compute as no simulations nor complex approximations are
required to obtain the critical value. Note that $\Sigma$ used in
the test procedures is the known covariance matrix associated to
the known profile specified by the null hypothesis.

\section{Example. Biological interpretation of a list of genes}

$CD4+$ T--cells are a type of white-blood cells which are very
important in the organism immune surveillance. As an example of
the many processes in which they are involved it is known that the
decrease in number of $CD4+$ T--cells is the primary mechanism by
which HIV causes AIDS. The activation of T--cells is related to
the presence of a substance, L-selectin ($CD62L$). This molecule
may be absent or present in a cell yielding two possible types of
cells: $CD4+/CD62L-$ T--cells lack L--selectin expression, whereas
$CD4+/CD62L+$ T--cells present L--selectin expression.

\cite{Hengel:2003} performed a study aiming at
finding the genetic regulation of these differences. They found
144 genes to be up--regulated in $CD4+/CD62L-$ T--cells relative
to $CD4+/CD62L+$ T--cells. The list is available in the
supplementary material website. All the computations in this work
have used only 140 of these genes, because there are 4 which were
not annotated in the GO.

Table \ref{TabProfiles} shows the functional profiles for this
list of genes at the first level of the three ontologies.

\begin{figure}[htbp]
\begin{center}
\includegraphics[width=\textwidth]{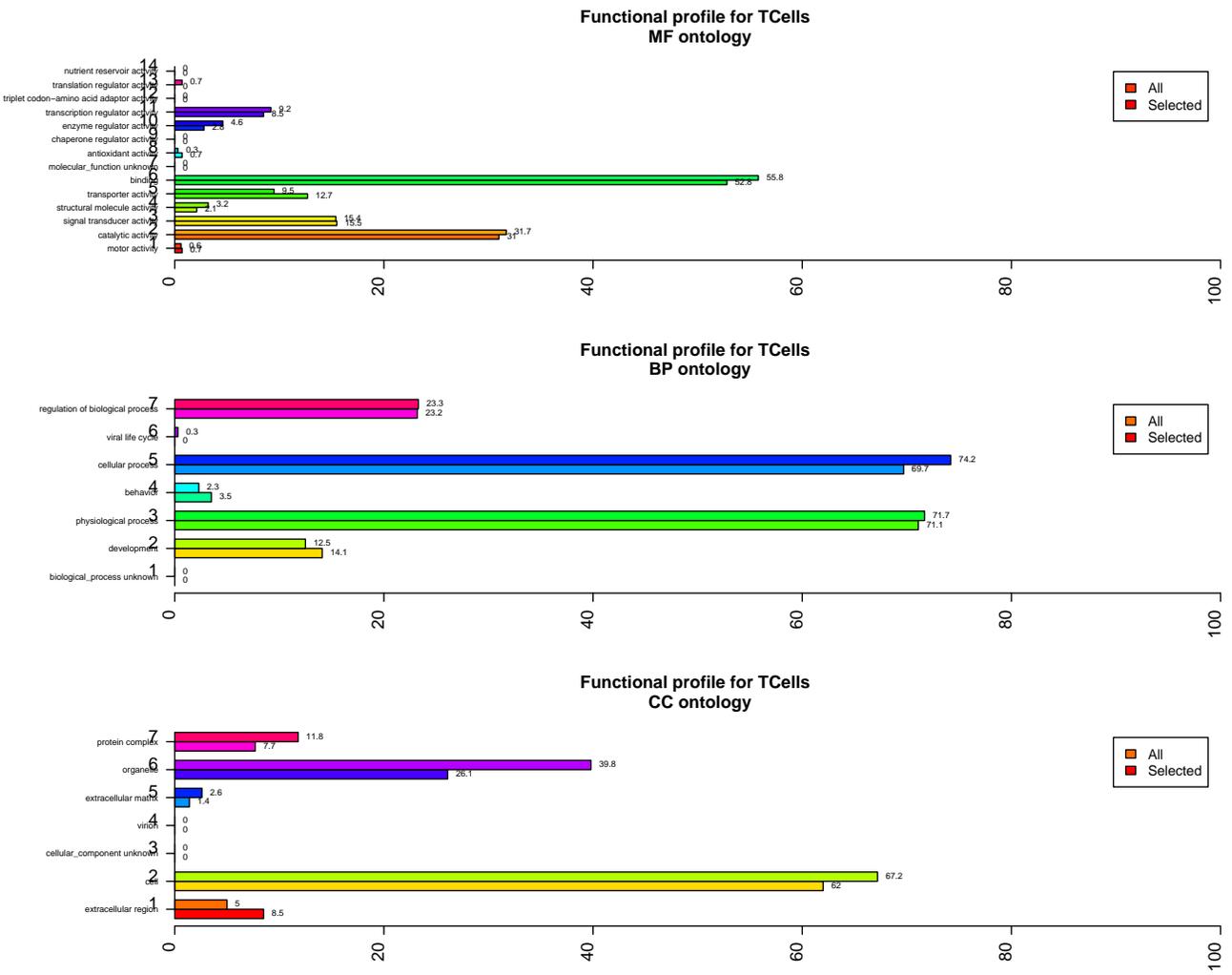}
\caption{comparison of GO profiles for all three ontologies at level
one of these 140 genes relative to all the genes in the array
(hgu95A)} \label{TCellsProfiles}
\end{center}
\end{figure}

There are two reasons why these profiles are not necessarily very
informative. First, for simplicity, we have built the profile at
the highest possible level formed by very generic groups. Even if
these functionalities are too general one might rely on them for
interpretation, but first we need to know if this profile
characterizes the selected genes or it is simply the profile
corresponding to the population, or in this case, to a big
sub--population formed by all the genes in the microarray. In
order to answer this second question a comparison of profiles is
meaningful. Figure \ref{TCellsProfiles} and table
\ref{TabProfileComp} show the comparison between the sample (140
genes in the list) and the population (all the genes on the chip)
profiles. It can be seen that there does not appear to be any
significant difference in BP and MF profiles, but there is some
for Cellular Component.

Table ~\ref{TabProfileComp} shows the distances and p--values
computed using the two methods described above. The test performed
here has null hypothesis $H_0: \dPPO=0$, that is, not rejecting
the null hypothesis means that the set of genes considered to be
differentially expressed is distributed between GO categories in
the same form as all the genes in the array.

\begin{table}[htbp]
\begin{center}
\begin{tabular}{|llr|}
\hline
{\bf Description} & {\bf GOID} & {\bf Frequency} \\
\hline
catalytic activity & GO:0003824 &         38 \\
signal transducer activity & GO:0004871 &         21 \\
structural molecule activity & GO:0005198 &          2 \\
transporter activity & GO:0005215 &         16 \\
binding & GO:0005488 &         76 \\
antioxidant activity & GO:0016209 &          1 \\
enzyme regulator activity & GO:0030234 &          4 \\
transcription regulator activity & GO:0030528 &         10 \\
\hline
{\bf Total number of hits} &     {\bf } &        168 \\
\hline
\end{tabular}
\end{center}
\caption{Functional profiles at the first level of the Molecular
Function Ontology for the list of genes in the example
data}\label{TabProfiles}
\end{table}

\begin{table}[h]
\begin{center}
\begin{tabular}{|clll|}
\hline
{\it Ontology} & {\it Distance} $\pm$ precision &{\it LC-chi p-value} & {\it Approx-chi p-value} \\
\hline
  {\bf MF} &  0.00734  $\pm$ 0.013114& 0.5744 & 0.6062\\
  {\bf BP} &  0.00511  $\pm$ 0.009710& 0.4112 & 0.4611\\
  {\bf CC} &  0.05749  $\pm$ 0.053666& 0.0000 & 3.11e-07\\
\hline
\end{tabular}

\end{center}
\caption{Distances and p--values computed on first--level profiles
for the three ontologies for the example data. ``LC-chi'' stands
for the test based on linear combinations of chi--square random
variables (\ref{RejectionBychiLinComb}) and ``Approx--chi'' stands
for the test based on approximating a chi--square distribution
(\ref{RejectionBychirank}) }\label{TabProfileComp}
\end{table}

\section{Simulation studies}

The accuracy of the preceding results and methods has been tested
by simulation. The simulated scenarios emulate the case of
$CD4+/CD62L+$ T--cells discussed in previous sections. Each
simulation was characterized by three main parameters: a sample
size $n$ and two expanded profiles $\eProf$ and $\eProf_0$ which
induced the profiles $\Prof= C \eProf$ and $\Prof_0=C \eProf_0$.

For each one of the simulation scenarios characterized by a given
combination of the above parameters, 10,000 (expanded) sample
profiles were generated according to a multinomial distribution
$\mathbf{M}(\eProf;n)$ and contracted according to
(\ref{contraction}). For each generated profile, the test
procedures (\ref{RejectionBychiLinComb}) and
(\ref{RejectionBychirank}) were applied in order to determine
whether the null hypothesis in (\ref{oneSampleTest}) was rejected
or not, and the confidence interval (18) computed to determine its
length and to inspect its coverage of the true distance $d\left(
{{\rm {\bf P}},{\rm {\bf P}}_0 } \right)$. These simulation
results were collected to estimate the true rejection
probabilities, the mean interval length and the true coverage
probability.

In a first series of simulation experiments, the profiles
specified in the preceding example were taken as directly defining
the population and/or the null hypothesis, with $n$ = 140.

Table ~\ref{simulationResults} displays the (simulation estimated)
probability of rejecting $H_{0}$, the mean length of the
confidence interval and its coverage. All results correspond to a
nominal significance level of 0.05 or to a nominal coverage of
0.95.

\begin{table}[htbp]
\begin{tabular}{|l|l|l|l|l|l|}
\hline
 \multicolumn{4}{|l}{} & \multicolumn{ 2}{|c|}{\it Profile specifying $H_0$} \\
\hline
 \multicolumn{2}{|l}{} &            &       Onto & {\it CD62L} & {\it hgu95A} \\
\hline
\multirow{24}{*}{Simulated profiles} & \multirow{12}{*}{{\it CD62L}} & \multicolumn{ 1}{|c|}{} &         BP & 0.0483$\pm$0.0042 & \multicolumn{ 1}{|c|}{} \\
&  & \multicolumn{ 1}{|c|}{LC--chi} &         MF & 0.0439$\pm$0.0040 & \multicolumn{ 1}{|c|}{} \\
&  & \multicolumn{ 1}{|c|}{} &  CC &
0.0475$\pm$0.0042 &
\multicolumn{ 1}{|c|}{} \\
&  & \multicolumn{ 1}{|c|}{} &         BP & 0.0529$\pm$0.0044 & \multicolumn{ 1}{|c|}{} \\
&  & \multicolumn{ 1}{|c|}{Approx--chi} &         MF & 0.0524$\pm$0.0044 & \multicolumn{ 1}{|c|}{} \\
&  & \multicolumn{ 1}{|c|}{} &         CC & 0.0475$\pm$0.0042 & \multicolumn{ 1}{|c|}{} \\
&  & \multicolumn{ 1}{|c|}{} &         BP & 0.0178$\pm$7.58E-5 & \multicolumn{ 1}{|c|}{} \\
&  & \multicolumn{ 1}{|c|}{Interval length} &         MF & 0.0220$\pm$8.59E-5 & \multicolumn{ 1}{|c|}{} \\
&  & \multicolumn{ 1}{|c|}{} &         CC & 0.0092$\pm$7.01E-5 & \multicolumn{ 1}{|c|}{} \\
&  & \multicolumn{ 1}{|c|}{} &         BP & 0.9951$\pm$0.0014 & \multicolumn{ 1}{|c|}{} \\
&  & \multicolumn{ 1}{|c|}{Coverage} &         MF & 0.9883$\pm$0.0021 & \multicolumn{ 1}{|c|}{} \\
&  & \multicolumn{ 1}{|c|}{} &         CC & 0.9986$\pm$0.0007 & \multicolumn{ 1}{|c|}{} \\
& \multirow{12}{*}{{\it hgu95A}} & \multicolumn{ 1}{|c|}{} &         BP & 0.1397$\pm$0.0068 & 0.0527$\pm$0.0044 \\
&  & \multicolumn{ 1}{|c|}{LC--chi} &         MF & 0.3555$\pm$0.0094 & 0.0550$\pm$0.0045 \\
&  & \multicolumn{ 1}{|c|}{} &         CC & 0.9057$\pm$0.0057 & 0.0453$\pm$0.0041 \\
&  & \multicolumn{ 1}{|c|}{} &         BP & 0.1455$\pm$0.0069 & 0.0557$\pm$0.0045 \\
&  & \multicolumn{ 1}{|c|}{Approx--chi} &         MF & 0.3692$\pm$0.0095 & 0.0574$\pm$0.0046 \\
&  & \multicolumn{ 1}{|c|}{} &         CC & 0.9198$\pm$0.0053 & 0.0518$\pm$0.0043 \\
&  & \multicolumn{ 1}{|c|}{} &         BP & 0.0215 $\pm$8.63E-5 & 0.0183$\pm$7.89E-5 \\
&  & \multicolumn{ 1}{|c|}{Interval length} &         MF & 0.0292$\pm$8.97E-5 & 0.0215$\pm$ 8.65E-5 \\
&  & \multicolumn{ 1}{|c|}{} &         CC & 0.0313$\pm$ 0.0001 & 0.0042$\pm$3.37E-5 \\
&  & \multicolumn{ 1}{|c|}{} &         BP & 0.9907$\pm$0.0019 & 0.9943$\pm$0.0015 \\
&  & \multicolumn{ 1}{|c|}{Coverage} &         MF & 0.9794$\pm$0.0028 & 0.9868$\pm$0.0022 \\
&  & \multicolumn{ 1}{|c|}{} &         CC & 0.8896$\pm$0.0061 & 0.9842$\pm$0.0024 \\
\hline
\end{tabular}
\caption{\label{simulationResults}Probability of rejecting $H_0$,
mean length of the confidence interval for the true squared
Euclidean distance and coverage. Simulated profiles were generated
according to \emph{CD62L} or \emph{hgu95A}. All simulation results
are displayed with $\pm$95\% confidence limits. ``LC--chi'' stands
for the test based on linear combinations of chi--square random
variables (\ref{RejectionBychiLinComb}) and ``Approx--chi'' stands
for the test based on approximating a chi--square distribution
(\ref{RejectionBychirank}) }
\end{table}

$H_{0}$ is true when the profile generating the gene samples and
the profile specifying $H_{0}$ are the same. In these cases, both
tests seem to perform according to the nominal significance level,
with an apparent tendency of test (\ref{RejectionBychirank}) to
slightly higher type I error probabilities. As can be expected,
the confidence interval (\ref{ConfidenceInt}) is not adequate when
$H_{0}$ is true, with a greater coverage than the nominal one and
a low precision (the intervals are too long in mean). When
$\Prof_0$ corresponds to \textit{CD62L} and $\Prof$ to
\textit{hgu95A}, $H_{0}$ is not true, though both profiles are
very similar in the BP and MF ontologies, with ``true'' squared
Euclidean distances of 0.0020 and 0.0064, respectively. For $n$ =
140, the power of the tests is low, around 0.14 for BP and 0.36
for MF. That is, for these quite similar ``true'' profiles, the
probability of type II error is high and the confidence interval
still performs not adequately. On the other hand, the simulated
profiles differ appreciably more for the CC ontology. Then the
power of the tests is much greater, around 0.91, but the coverage
is still inadequate, now too low.

In order to have a more comprehensive view, we performed an
additional simulation study. The following geometric model was
considered:

Let $\eProf = \left( {p_{11} ,p_{22} ,\ldots ,p_{ss} ,p_{12}
,p_{13} ,\ldots,p_{(s-1)s} ,p_{123} ,\ldots } \right)^t$ represent
an expanded profile. Maintaining the order of the elements, recode
the indexes as
\[
\left( {p_{[1]} ,p_{[2]} ,\ldots ,p_{[s]} ,p_{[s + 1]} ,p_{[s + 2]}
,\ldots ,p_{\left[ {\frac{s\left( {s + 1} \right)}{2}} \right]}
,p_{\left[ {\frac{s\left( {s + 1} \right)}{2} + 1} \right]} ,\ldots
} \right)
\]
and assume that
\begin{equation}
\label{eq1} p_{[i]} \propto \left( {1 - \theta } \right)^{i - 1},
\mbox{ for } 0 < \theta < 1.
\end{equation}

The sole purpose of model (\ref{eq1}) is to define families of
profiles fully characterized by a unique parameter $\theta$, for a
given number of ontology classes $s$ and a given maximum level of
possible simultaneity $k$ --that is, with the last term having
index $\left( {s - k + 1} \right)\left( {s - k + 2} \right)\ldots
\left( {s - 1} \right)s$. Progressively different values of
$\theta$ produce progressively greater distances between profiles,
that is, scenarios progressively distant from the assumption of
validity of the null hypothesis. In all the simulations, $H_{0}$
was defined by a profile associated to $\theta_{0}$ = 0.4, while
$\theta$ = 0.4, 0.35, 0.30, 0.25 and 0.20 defined possible
scenarios were the sample profiles were generated according to
distributions more and more distant from $H_{0}$. Additionally,
the following sample sizes (number of genes) were considered: $n$
= 50, 100, 140, 200, 500 and 1000. Here we report the results for
$s$ = 6 and $k$ = 5 (as in the $CD4+/CD62L$ T-cells example for
the BP ontology) for a nominal significance level of 0.05 and
nominal coverage of 0.95. The results of other situations
(including $s$ = 11, $k$ = 6 and $s$ = 4, $k$ = 2, respectively
the case of the MF and CC ontologies) are accessible in the
supplementary documentation web site.

\begin{figure}[htbp]
\centerline{\label{PowerOfChiApproxTest}\includegraphics{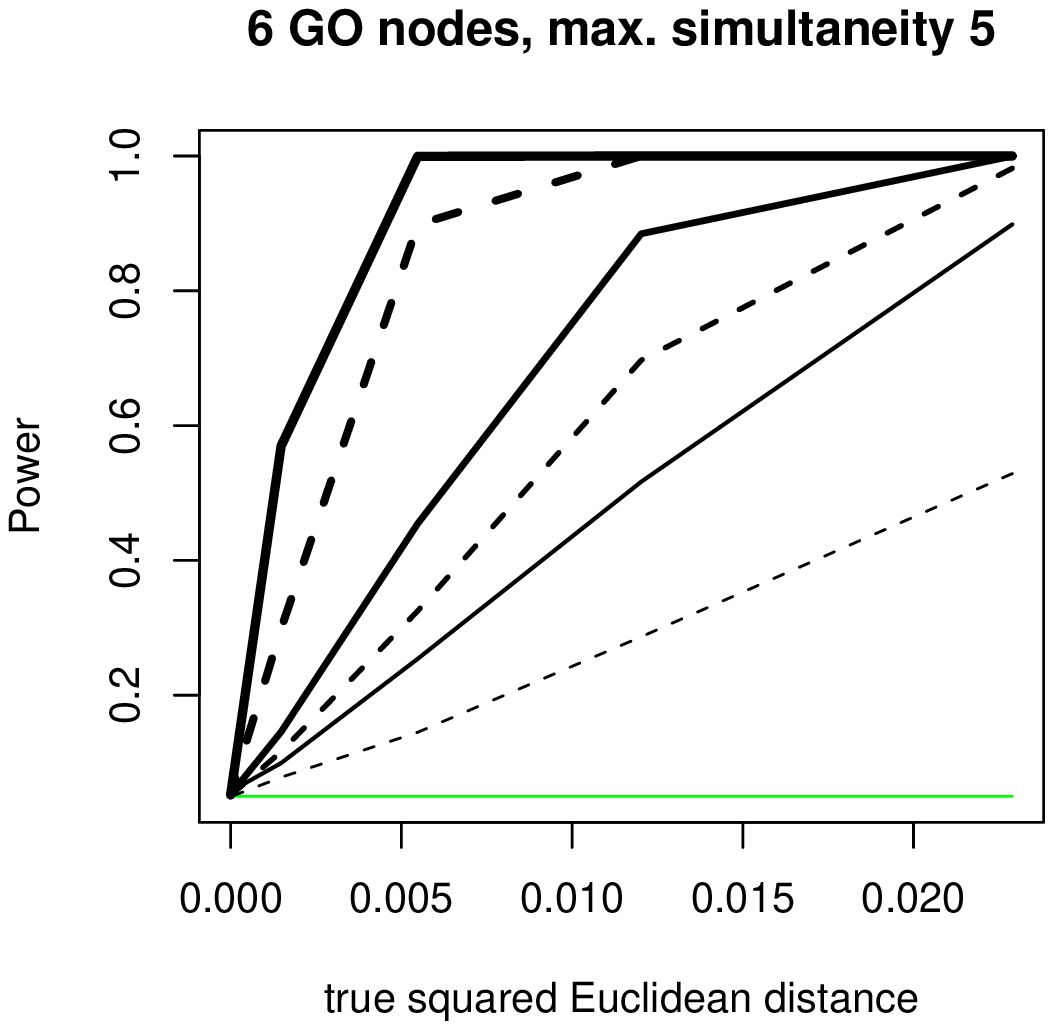}}
\caption{Power of the approximated chi-square test. The bottom
horizontal line corresponds to the reference 0.05 significance
level.}
\end{figure}

Figures \ref{PowerOfChiCLTest} and \ref{PowerOfChiApproxTest}
display the power curves of both tests under consideration. They
perform in a very similar way and always seem to be in conformity
with the nominal significance level. Thus, the test based on the
chi-square approximation seems to be preferable due to its
simplicity.

\begin{figure}[htbp]
\centerline{\includegraphics{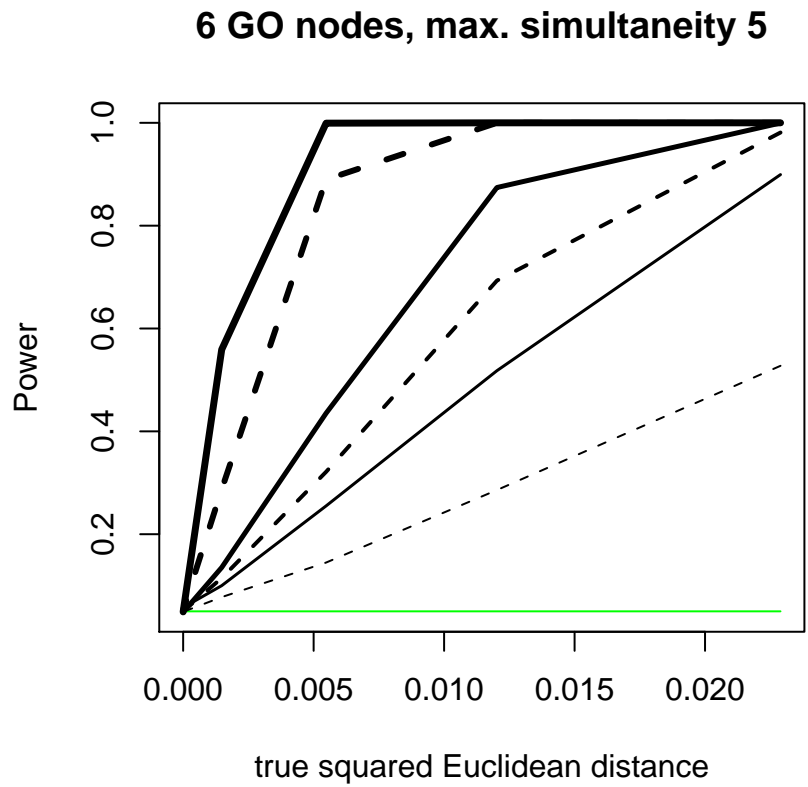}}
\caption{\label{PowerOfChiCLTest}Power of the test based on linear
combinations of chi--square random variables. The bottom
horizontal line corresponds to the reference 0.05 significance
level.}
\end{figure}

\begin{figure}[htbp]
\centerline{\includegraphics{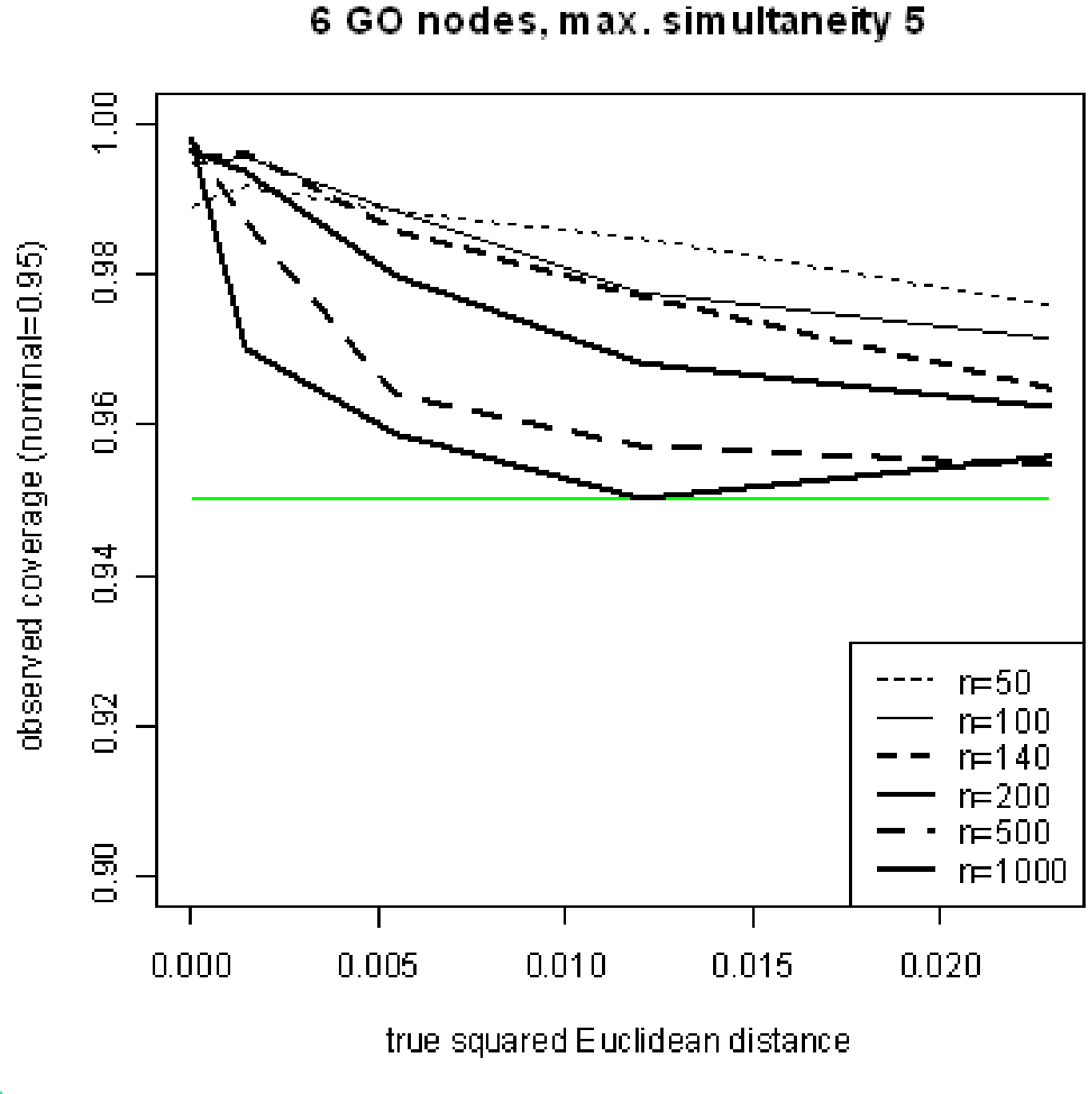}}
\caption{\label{CoverageOfCI}Coverage of the asymptotic confidence
interval. The base reference line corresponds to a 0.95 coverage.}
\end{figure}

\begin{figure}[htbp]
\centerline{\includegraphics{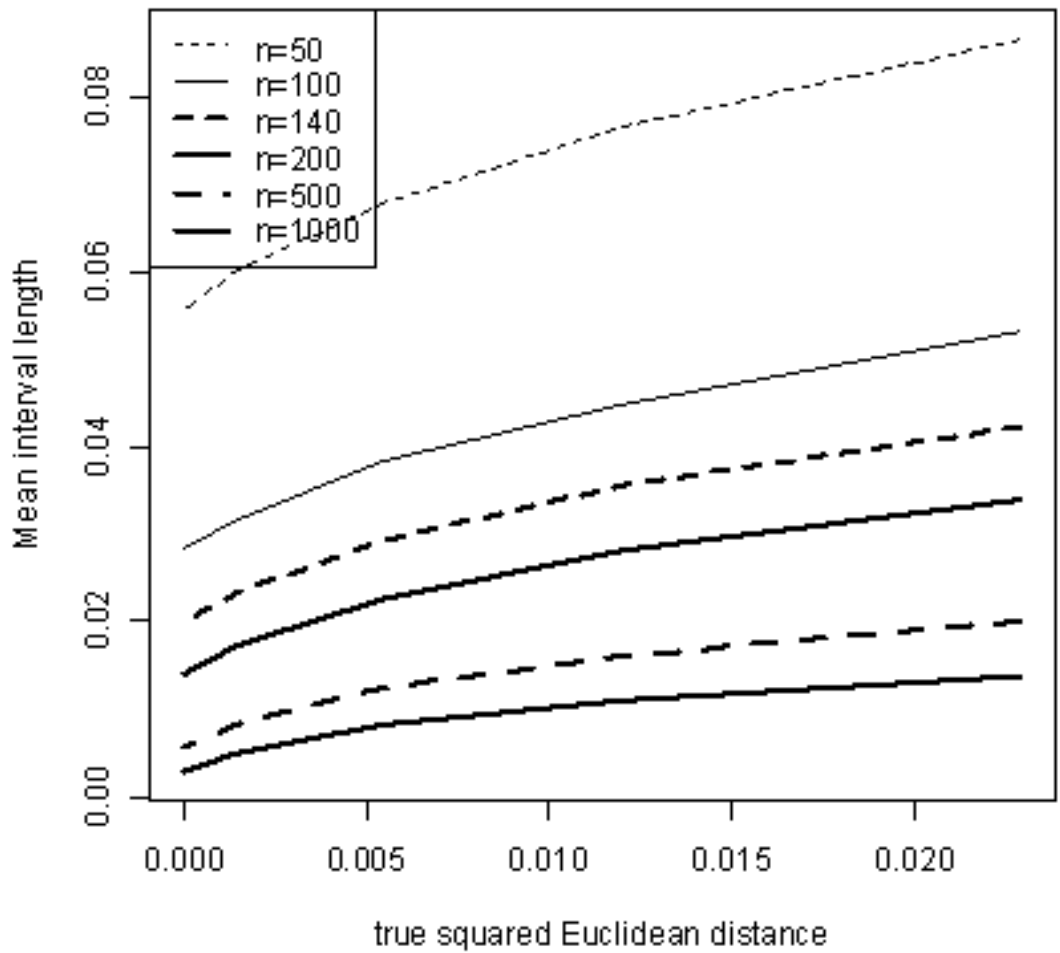}}
\caption{\label{MeanCILength}Mean confidence interval length}
\end{figure}

Figure \ref{CoverageOfCI} corresponds the coverage of the
asymptotic confidence interval (\ref{ConfidenceInt}). As is
expected, this confidence interval is not adequate under true
$H_{0}$. When $H_{0}$ is false, only for large sample sizes (500
or more genes) its true coverage approximates the nominal one.
Otherwise the true coverage is larger than the nominal, but at the
cost of a very low precision (that is, too long intervals) as is
shown in Figure \ref{MeanCILength}. For example, if $n$ = 50
genes, with a true 0.005 distance, a (nominally) 95{\%} confidence
interval will have a length of approximately 0.04, too wide with
respect to the magnitude of the distance.

\clearpage

\section{Discussion and Conclusion}

The analysis and interpretation of biological data based on the
Gene Ontology is an active field of research.

Functional profiles constitute an intuitive way to summarize
\emph{sets} of genes of any size to facilitate biological
interpretation.

Our theoretical results set the basis for doing statistical
inference based on these profiles. This allows to turn the
analysis of profiles from a simple graphical comparison, such as
is done in many papers (\cite{Beltran:2003}) or in the
\emph{goTools} Bioconductor package (\cite{Paquet:2005})-- to well
based inferential procedures with a known degree of confidence.

Essentially we have considered \emph{global comparisons} between
profiles at fixed levels of the GO, but extensions are
straightforward. For instance, profiling may be performed on any
set of reference categories, not necessarily a fixed level. Also,
the theory can be easily adapted to other situations such as the
analysis of multiple response items in surveys.

The approach is, of course, not free from limitations. Profiling,
as any other summary, implies a certain loose of information.
Comparing the approach with the use of the whole graph it is clear
that the later has more information but is more difficult to
summarize. If we go in the other direction, a category by category
analysis (``a la fatiGO'') helps to see what happens in specific
interesting categories but does not offer a global approach. In
brief our approach tries to stay between one and the other in a
useful way.

\subsection{Software and tools}

The main applied interest of this work is to provide the genomic
community a research tool that help to assess their conclusions,
allowing to go one step further than visual approximations such as
that offered by some programs, such as the Bioconductor package
\Rpackage{goTools} (\cite{Paquet:2005}).

To facilitate the application of our results we have developed a
tool which is available as an \R\ package which will be freely
accessible to the user's community. This will also be submitted to
Bioconductor to help its diffusion. Also, a web site to make the
software available through the web is in development.

\section*{Acknowledegments} We are grateful to Dr. Sandrine Dudoit,
at U.C. Berkeley, for her comments and support during Dr. Alex
S\'anchez's stay in Berkeley, where this work was initiated.

\bibliography{Statistics4FunctionalProfiles}

\begin{thebibliography}{15}
\expandafter\ifx\csname natexlab\endcsname\relax\def\natexlab#1{#1}\fi
\expandafter\ifx\csname url\endcsname\relax
  \def\url#1{\texttt{#1}}\fi
\expandafter\ifx\csname urlprefix\endcsname\relax\def\urlprefix{URL }\fi

\bibitem[{Al-Shahrour et~al.(2004)Al-Shahrour, Diaz-Uriarte, and
  Dopazo}]{fatiGO:2004}
Al-Shahrour, F., Diaz-Uriarte, R., Dopazo, J., 2004. \texttt{FatiGO}: a web
  tool for finding significant associations of gene ontology terms with groups
  of genes. Bioinformatics 20, 578--580.

\bibitem[{Alizadeh et~al.(2000)Alizadeh, Eisen, Ma, , Lossos, Rosenwald,
  Boldrick, Sabet, Tran, Yu, Powell, Yang, Marti, Jr, Lu, Lewis, Tibshirani,
  Sherlock, Chan, Greiner, Weisenburger, Armitage, Warnke, Levy, Wilson,
  Grever, Byrd, Botstein, Brown, and Staudt}]{Alizadeh:2000}
Alizadeh, A., Eisen, M., Ma, E. D.~C., , Lossos, I., Rosenwald, A., Boldrick,
  J., Sabet, H., Tran, T., Yu, X., Powell, J., Yang, L., Marti, G., Jr, J.~H.,
  Lu, L., Lewis, D., Tibshirani, R., Sherlock, G., Chan, W., Greiner, T.,
  Weisenburger, D., Armitage, J., Warnke, R., Levy, R., Wilson, W., Grever, M.,
  Byrd, J., Botstein, D., Brown, P., Staudt, L., February 2000. Distinct types
  of diffuse large {B}--cell lymphoma identified by gene expression profiling.
  Nature 403, 503--511.

\bibitem[{Alon et~al.(1999)Alon, Barkai, Notterman, Gish, Ybarra, Mack, and
  Levine}]{Alon:1999}
Alon, U., Barkai, N., Notterman, D., Gish, K., Ybarra, S., Mack, D., Levine,
  A., 1999. Broad patterns of gene expression revealed by clustering analysis
  of tumor and normal colon tissues probed by oligonucleotide arrays.
  Proceedings of the National Academy of Sciences 96, 6745--6750.

\bibitem[{Beltran et~al.(2003)Beltran, Blanco, Serra, Perez-Villamil,
  Guig{\'o}, Artavanis-Tsakonas, and Corominas}]{Beltran:2003}
Beltran, S., Blanco, E., Serra, F., Perez-Villamil, B., Guig{\'o}, R.,
  Artavanis-Tsakonas, S., Corominas, M., 2003. Transcriptional network
  controlled by the trithorax- group gene ash2 in drosophila melanogaster
  genome research. Proceedings of the National Academy of Sciences 100,
  3293--3298.

\bibitem[{Dennis et~al.(2003)Dennis, Sherman, Hosack, Yang, Baseler,
  Clifford~Lane, and Lempicki}]{DAVID:2003}
Dennis, G.~J., Sherman, B., Hosack, D., Yang, J., Baseler, M.~W.,
  Clifford~Lane, H., Lempicki, R., 2003. David: Database for annotation,
  visualization, and integrated discovery. Genome Biology 4, P3.

\bibitem[{Dik and Gunst(1985)}]{Dik:1985}
Dik, J., Gunst, M., 1985. The distribution of general quadratic forms in normal
  variables. Statistica Neerlandica 39, 14--26.

\bibitem[{Farebrother(1984)}]{Farebrother:1984}
Farebrother, R., 1984. The distribution of linear combinations of $\chi^2$
  random variables. Appl. Statist. 33, 366--369.

\bibitem[{Hengel et~al.(2003)Hengel, Thaker, Pavlick, Metcalf, Dennis, Yang,
  Lempicki, Sereti, and Lane}]{Hengel:2003}
Hengel, R., Thaker, V., Pavlick, M., Metcalf, J., Dennis, G.~J., Yang, J.,
  Lempicki, R., Sereti, I., Lane, H., 2003. L-selectin (cd62l) expression
  distinguishes small resting memory cd4+ t cells that preferentially respond
  to recall antigen. J. Immunol. 170, 28--32.

\bibitem[{Mootha et~al.(2003)Mootha, Lindgren, Eriksson, Subramanian, Sihag,
  Lehar, Puigserver, Carlsson, Ridderstrale, Laurila, Houstis, Daly, Patterson,
  Mesirov, Golub, Tamayo, Spiegelman, Lander, Hirschhorn, Altshuler, and
  Groop}]{Mootha:2003}
Mootha, V., Lindgren, C., Eriksson, K., Subramanian, A., Sihag, S., Lehar, J.,
  Puigserver, P., Carlsson, E., Ridderstrale, M., Laurila, E., Houstis, N.,
  Daly, M., Patterson, N., Mesirov, J., Golub, T., Tamayo, P., Spiegelman, B.,
  Lander, E., Hirschhorn, J., Altshuler, D., Groop, L., 2003.
  Pgc-1alpha-responsive genes involved in oxidative phosphorylation are
  coordinately downregulated in human diabetes. Nature Genet. 34, 267--73.

\bibitem[{Mosquera and S{\'a}nchez-Pla(2005)}]{Mosquera:2005}
Mosquera, J.-L., S{\'a}nchez-Pla, A., 2005. A comparative study of go mining
  programs. In: X Conferencia Espa{\~n}ola de Biometr{\'i}a. Sociedad
  Espa{\~n}ola de Biometr{\'i}a.

\bibitem[{Paquet and Yang(2005)}]{Paquet:2005}
Paquet, A., Yang, Y., 2005. Getting started with \texttt{goTools} package.
\newline\urlprefix\url{http://bioconductor.org/repository/devel/vignette/goToo%
ls.pdf}

\bibitem[{Rao and Scott(1981)}]{raoscott:1981}
Rao, J., Scott, A., 1981. The analysis of categorical data from complex sample
  surveys: chi-squared tests for goodness of fit and independence in two way
  tables. JASA 76, 221--3.

\bibitem[{Serfling(1980)}]{Serfling:1980}
Serfling, R., 1980. Approximation Theorems of Mathematical Statistics. John
  Wiley, New York.

\bibitem[{Sheil and O'Muir\-cheartaigh(1977)}]{Sheil:1977}
Sheil, J., O'Muir\-cheartaigh, I., 1977. Algorithm \texttt{AS}106. \textsc{T}he
  distribution of non-negative quadratic forms in normal variables. Appl.
  Statist. 26, 92--98.

\bibitem[{Smyth(2005)}]{Smyth:2005}
Smyth, G., 2005. Bioinformatics and Computational Biology Solutions using R and
  Bioconductor. Springer, New York, Ch. Limma: linear models for microarray
  data, pp. 397--420.

\end{thebibliography}

\end{document}